\documentclass[journal,transmag]{IEEEtran}
\usepackage{cite}
\usepackage{bm}
\usepackage{amsmath,amssymb, amsthm}
\usepackage{tabularx}
\usepackage{multirow}
\usepackage{graphicx}
\usepackage{latexsym}
\usepackage{color}
\usepackage{caption}
\usepackage[table]{xcolor}
\usepackage{subcaption}
\usepackage{bm}
\usepackage{bbm}
\usepackage{amsmath,amssymb, amsthm}
\usepackage{algorithm}

\usepackage{algpseudocode}
\usepackage{color,soul}
\usepackage{graphics}
\usepackage{amsfonts}
\usepackage{setspace}   
\usepackage{wasysym}
\usepackage{multirow}

\usepackage{adjustbox}
\usepackage{graphicx}
\usepackage{multirow}

\usepackage{nomencl}
\usepackage{float}
\begin{document}
\title{Efficient Collection of Connected Vehicles Data with Precision Guarantees  }


\author{\IEEEauthorblockN{Negin Alemazkoor\IEEEauthorrefmark{1},
Hadi Meidani*\IEEEauthorrefmark{1}}
\IEEEauthorblockA{\IEEEauthorrefmark{1}Department of Civil and Environmental Engineering,
	University of Illinois at Urbana-€"Champaign, Urbana, IL 61801 USA}
\thanks{ 
Corresponding author: Hadi Meidani (email: meidani@illinois.edu).}}

\IEEEtitleabstractindextext{%
\begin{abstract}
Connected vehicles disseminate detailed data, including their position and speed, at a very high frequency. Such data can be used for accurate real-time analysis, prediction and control of transportation systems. The outstanding challenge for such analysis is how to continuously collect and process extremely large volumes of data. To address this challenge, efficient collection of data is critical to prevent overburdening the communication systems and overreaching computational and memory capacity. In this work, we propose an efficient data collection scheme that selects and transmits only a small subset of data to alleviate data transmission burden. As a demonstration, we have used the proposed approach to select data points to be transmitted from 10,000 connected vehicles trips available in the Safety Pilot Model Deployment dataset. The presented results show that collection ratio can be as small as 0.05 depending on the required precision. Moreover, a simulation study was performed to evaluate the travel time estimation accuracy using the proposed data collection approach. Results show that the proposed data collection approach can significantly improve travel time estimation accuracy. 
\end{abstract}

}

\maketitle

\section{Introduction}
Intelligent transportation systems (ITS) are established to actively manage traffic in order to improve mobility, efficiency, reliability and safety \cite{wang2010parallel,dimitrakopoulos2010intelligent}. This is typically done by using  real-time traffic data information collected from various sources such as traffic sensors and smart phones \cite{mitrovic2015low}. With increasing penetration of connected vehicles into transportation systems, highly detailed data from connected vehicles can provide great opportunity for real-time analysis toward optimal system management. However, collecting such detailed data with high frequency from several thousands of vehicles can easily lead to data flood or data explosion, where data communication and storage capacities are overreached. For instance, if for each vehicle we only collect the coordinates every 10 seconds, the storage required for 400 vehicles in a single day will exceed 100 megabytes \cite{meratnia2004spatiotemporal}. This is while the connected vehicles transmit data at an even higher frequency, i.e. every 0.1 second, and they also  transmit other information, including but not limited to speed. In \cite{WinNT}, the data uploaded to the cloud from one connected vehicle in an hour is reported to be about 25 gigabytes. In \cite{mai2011business} it is estimated that 400 million gigabytes of data would be transferred each month, if 25 percent of all vehicles are connected. Therefore, efficient data collection and compression is vital to facilitate handling such large volume of data. 

There are two main types of data compression methods: lossless and lossy. In lossless compression, original data is restored without any loss when the file is uncompressed. Well-known examples of lossless compression are WinZip \cite{WinZip} and WinRAR \cite{roshal2009winrar}. Lossless compression methods are used when even a slight difference between original and reconstructed files is not acceptable. For instance, text files and source codes typically require lossless compression \cite{witten1999text}. On the other hand, lossy compression techniques incur inaccuracies and loss of information. Therefore, exact recovery of a file compressed using a lossy method is not possible. However, in return, lossy methods enable significantly higher compression ratios compared to lossless methods \cite{sayood2017introduction}, and effectively offer a trade-off between the compression ratio and acceptable information distortion. 

Several works have used lossy approaches to compress traffic data obtained from traffic sensors. These approaches mostly compress signals by decorrelating data points and transforming them into a new domain. In \cite{li2007flow}, principal component analysis (PCA) was used to compress traffic flow data. It was shown that 87\% recovery accuracy can be achieved by 6.2\% compression ratio. In \cite{ding2011method}, a wavelet-PCA method based was proposed to compress urban traffic data. In this method, first a wavelet decomposition of data is acquired, to which PCA is then applied  to further reduce the dimensionality of data. It was shown that the wavelet-PCA method outperforms the conventional PCA approach in terms of compression ratio and accuracy. Several other works have also used wavelet transformation for traffic data compression \cite{agarwal2017multidimensional, xiao2004traffic,cheng2007mining,qiao2006incorporating}. It should be noted that in all these  methods, it is required to collect all the data points. Therefore, even though these approaches alleviate the  \emph{storage} burden, they do not address  potential  data \emph{transmission} bottlenecks. It is thus vital to also avoid overwhelming communication systems by wisely selecting the data points that are to be collected, thereby minimizing the number data transmissions.

 In \cite{lin2018efficient}, for the first time, a compressive sampling approach was proposed for efficient collection of connected vehicles data. The authors showed that with the collection ratio of 0.2, i.e. with collecting data at each time with 20\% chance, the original data can be recovered with 95\% accuracy using compressive sampling. In this work, we aim to achieve a high accuracy with a significantly smaller collection ratio. To this end, we propose an online multidimensional piecewise linear approximation (online-MPLA) approach for efficient connected vehicles data collection. To demonstrate the efficiency of proposed method, we compare the method with uniform and compressive sampling approaches. In summary, the contributions of this paper are the following:
 
 \quad 1. Proposing an online multidimensional piecewise linear approximation approach for efficient collection of connected vehicles data.
 
 \quad 2. Performing an extensive empirical study using real world connected vehicles data to show the effectiveness of proposed data collection method in comparison with other available methods.
 
\quad 3. Performing a simulation study to evaluate travel time estimation accuracy when online-MPLA is used for efficient data collection. 

 The rest of this paper is organized as follows. Section \ref{sec:Background} includes an overview on uniform and compressive sampling as well as discrete cosine transform, which is used in the compressive sampling technique. Section \ref{sec:Method} introduces our methodology for efficient data collection. Section \ref{sec:Exp} includes the empirical results and discussion. Section \ref{sec:Conclusion} highlights future research directions and concluding remarks.   
 \section{Technical background} \label{sec:Background}  
 In this section, we first introduce discrete cosine transformation as a lossy compression approach, and then review two available sampling approaches for efficient collection of connected vehicle data.  
\subsection{Discrete cosine transform} \label{sec:DCT}
Discrete cosine transformation (DCT) is a linear transform that maps a vector, i.e. a sequence of data points, into a summation of orthogonal cosine basis functions. DCT is similar to  discrete Fourier transform (DFT). The main difference between DCT and DFT is that DCT applies to real numbers, while DFT can be applied to complex numbers. Let $ \bm x  = [x_1, \cdots, x_N ] \in \mathbb{R}^N$ be a sequence of data points. A discrete cosine transform, $\bm \alpha$ is then defined as follows,
\begin{equation}\label{eq:dct}
\begin{small}
\alpha_i=\sqrt{\frac{2}{N}}\sum_{j=1}^{N}x_j \lambda_i \cos\left(\frac{\pi (j-0.5)(i-1)}{N}\right), \quad 1\leq i\leq N,
\end{small}
\end{equation}
where $\lambda_i= 1$ for $i>1$ and $\lambda_1= 1/\sqrt{2}$. Consequently, the inverse DCT is given by
 \begin{equation}\label{eq:idct}
 \begin{small}
 x_i=\sqrt{\frac{2}{N}}\sum_{j=1}^{N}\alpha_j \lambda_j \cos\left(\frac{\pi (j-1)(i-0.5)}{N}\right), \quad 1\leq i\leq N.
 \end{small} 
 \end{equation} 
 Equations (\ref{eq:dct}) and  (\ref{eq:idct}) define "DCT-II" that is the most commonly used DCT and is usually referred to as "the DCT" \cite{chen2005pca}. Let us define DCT matrix, $\bm \Psi$ as 
\begin{equation}
 \begin{small}
\bm \Psi_{ij}=\sqrt{\frac{2}{N}} \lambda_i \cos\left(\frac{\pi (j-0.5)(i-1)}{N}\right), \quad 1\leq i, j\leq N. 
 \end{small}
\end{equation}
Then  Equations (\ref{eq:dct}) and  (\ref{eq:idct}) can be rewritten as
\begin{equation}\label{eq:dct1}
\bm \alpha= \bm \Psi \bm x,
\end{equation}
\begin{equation}\label{eq:idct1}
\bm x= \bm \Psi^{-1} \bm \alpha.
\end{equation}
It should be noted that $\bm \Psi$ is a unitary matrix, hence $\bm \Psi^{-1}=\bm \Psi^{T}$. To compress data vector $\bm x$, first Equation (\ref{eq:dct1}) is used to obtain discrete cosine transformation, $\bm \alpha$. Then, $s$ largest entries of $\bm \alpha$ are kept and the rest are set to be zero, forming an approximate cosine transformation, $\hat{\bm \alpha}_s$, for storage purposes. Compression ratio depends on the $s$ value, which is set based on the acceptable data distortion. An approximation of original data, $\hat{\bm x}$, can be readily obtained by replacing $\hat{\bm \alpha}$ in Equation (\ref{eq:idct1}):
\begin{equation}\label{eq:idct1app}
	\hat{\bm x}= \bm \Psi^{T} \hat{\bm \alpha}_s.
\end{equation}
\subsection{Compressive sampling}
Compressive sampling  first appeared in the field of signal processing, with the objective of recovering a sparse signal with significantly smaller number of samples compared to that from the conventional Shannon-Nyquist sampling rate \cite{candes2008introduction}. Compressive sampling has been extensively applied in various fields where small sampling rate is desirable \cite{lustig2005application,ender2010compressive, gemmeke2010compressive,alemazkoor2018near}. In \cite{lin2018efficient}, motivated by the fact that discrete cosine transformation for connected vehicles data could be (approximately) sparse, compressive sampling was used to perform discrete cosine transformation with a small number of samples, thereby reducing the number of data transmissions. In particular, consider vector $\bm x^M$ to be the vector of  $M$ collected samples, i.e. $\bm x^M = \bm D \bm x $, where $\bm D$ is formed by randomly selecting $M$ rows of identity matrix of size $N$. Discrete cosine transformation can then be estimated using compressive sampling, as follows,
\begin{equation} \label{eq:l0min}
\underset{\bm \alpha}{\textrm{min}} \left \| \bm \alpha \right \|_{0} \quad \textrm{subject to} \quad   \bm x^M=\bm \Theta \bm \alpha,
\end{equation}
where $\bm \Theta= \bm D \bm \Psi^T$. Since the above $\ell_0$-norm minimization is NP-~hard, $\ell_0$ is usually replaced with its convex relaxation and $\ell_1$-norm is minimized instead, i.e.,
\begin{equation} \label{eq:l1min}
\underset{\bm \alpha}{\textrm{min}} \left \| \bm \alpha \right \|_{1} \quad \textrm{subject to} \quad   \bm x^M=\bm \Theta \bm \alpha.
\end{equation}
For accurate approximation of $\bm \alpha$ using (\ref{eq:l1min}), $\bm \alpha$ must be sufficiently sparse and $\bm \Theta$ must be incoherent. We refer interested readers to \cite{donoho2006compressed, candes2008restricted,candes2006robust} for theorems and more details regarding recovery accuracy associated with $\ell_1$-norm minimization. 

In \cite{lin2018efficient}, the speed data from Safety Pilot Model Deployment dataset was used to investigate the applicability of compressive sampling in efficient collection of connected vehicles speed data. It was shown that 95\% accuracy can be achieved on average with a collection ratio of 0.2. However, the applicability of compressive sampling was only shown for speed data. To be able to exploit collected speed data, collecting coordinates of connected vehicles is also necessary. In section \ref{sec:Exp}, we also report the trajectory approximation accuracy obtained by  compressive sampling. We compare recovery accuracy offered by compressing sampling with that offered by our proposed method in Section \ref{sec:Method} and also by a na\"{i}ve sampling approach, i.e. uniform sampling approach,  discussed in the next section. 
\subsection{Uniform sampling}
Uniform sampling can be thought of as the most na\"{i}ve method among lossy compression approaches\cite{muckell2011squish}. This approach uniformly subsamples streaming data. In other words, in uniform sampling, every $i$th data point (with an integer $i$) from the time series data is collected and stored. Sampled points can be connected with line segments to approximate the original signal. The main advantage of uniform sampling is its great simplicity, while its main  disadvantage is its inability to capture rapid changes that could take place between sampled points. 

In what follow, we introduce our proposed methodology for efficient collection of connected vehicle data.

\section{Methodology}\label{sec:Method}
In this section, we first review piecewise linear approximation (PLA), and then introduce our proposed methodology, which is an online multidimensional piecewise linear approximation for efficient collection of connected vehicle data. 
\subsection{Piecewise linear approximation}
PLA is a classic problem and has been widely used for time series data compression. PLA  approximates time series using a number of continuous or disjoints line segments. The two main measures for evaluating the quality of PLA are approximation size and error. There are three major classes of PLA algorithms: (a) bottom-up, (b) top-down, and (c) sliding window. In bottom-up algorithms, first, the finest approximation of time series with $N/2$ segments is considered, where $N$ is the length of time series. Then, these fine segments are merged iteratively until an stopping criteria is met \cite{hunter1999knowledge, heckbert1997survey}. On the contrary, in top-down algorithms, first, an approximation of time series with only two segments is considered. Then, these two segments are iteratively split until the specified approximation error tolerance is met \cite{douglas1973algorithms, park1999fast}. Sliding window algorithms consists in creating a segment at the first point of time series and expanding it to the right, by iteratively including new points in the segment. Once the segment exceeds acceptable approximated error tolerance at point $i$, the algorithm creates a new segment at that point \cite{koski1995syntactic, wang2000supporting,vullings1997ecg}. 
    
\subsection{Efficient connected vehicle data collection}
Bottom-up and top-down algorithms cannot be used for reducing the volume of data that connected vehicle transmit. This is because these algorithms are offline and require full collection of time series before running the algorithm. On the other hand, sliding window algorithms are online, meaning that the PLA algorithm can start at the same time as first data point collection. However, many variants of sliding window algorithms still need to collect all data points as an optimization problem must be solved each time a new data point is received in order to optimize the line segment to achieve the minimum approximation error \cite{elmeleegy2009online, luo2015piecewise}. 

The only two variants of sliding window that do not require collecting all the data points are cache filtering and linear filtering. A cache filter predicts the next incoming data point to have the same value as previous data point. If the prediction be within the specified error threshold, then the prediction is acceptable and the new data will not be recorded. In other words, a recording is only made when the prediction exceeds the specified error threshold \cite{olston2003adaptive}. In linear filtering, a line is passed through the first two points. This line is then used to make predictions for incoming new data points. Once the prediction error is larger than the desired error threshold, a new recording will be made and a new line segment will be created. Linear filters can approximate time series using jointed or disjointed line segments. In case of disjointed line segments, the new line segment is defined with the new recording and its next incoming data point. In case of jointed line segments, the new line segment is defined with the new recording and the last point of previous line segment \cite{dilman2002efficient}. In cache and linear filtering, the specified error threshold is in fact the $\ell_\infty$ error for the linear approximation. In other words, using these filters, it is guaranteed that approximation error is equal or smaller than the threshold at any point in the approximated time series. 

In this work, we propose a multidimensional linear filter with disjointed segments for efficient collection of connected vehicle data. We also impose a maximum segment length criteria, which constrains the maximum number of consecutive unrecorded data points. This is to account for  cases where the connection is impaired and the operation center is unaware of the situation, and trusting predictions obtained by linear segment could be misleading. It should be noted that there should be an agreement between individual connected vehicles and the  operation center on the type of data that must be transmitted, the error threshold, and maximum segment length.

Algorithm \ref{pseudocode-CV} summarizes the online-MPLA algorithm from a connected vehicle standpoint.  Let $d$ be the data dimension, i.e. the number of parameters that are recorded, for that vehicle at each time.  In this Pseudocode, $\bm X_t \in  \mathbb{R}^{1 \times d}$ and $\tilde{\bm X}_{t} \in  \mathbb{R}^{1 \times d}$ are exact and approximated information associated with the connected vehicle at time step $t$, $\bm \delta$ is the current segment's slope, and $L$ is the current segment's length.   
 \begin{algorithm}[H] 
 	\caption{Online-MPLA from connected vehicle standpoint}\label{pseudocode-CV}
 	\begin{algorithmic}[1]
 		\State Set type of data that must be transmitted.
 		\State Set the error threshold, $\bm \epsilon=(\epsilon^1,\cdots,\epsilon^d).$ 
 		\State Set maximum segment length, $K$.
 		\State Transmit $\bm X_{1}$ \& $\bm X_{2}$.
 		\State Initialize $\bm \delta=\bm X_{2}-\bm X_{1}$ \& $L=2$.
        \While {Trip has not ended}
       \State $\tilde{\bm X}_{t}$=$\tilde{\bm X}_{t-1}+\bm \delta$
        \If {$|\tilde{\bm X}_{t}^i-\bm X_{t}^i|>\epsilon^i$ for any $1\leq i\leq d$ or $L>K$}
        \State{Transmit $ \bm X_{t}$ \& $ \bm X_{t+1}$},
        \State Set $\tilde{\bm X}_{t}={\bm X}_{t} $ \& $\tilde{\bm X}_{t+1}={\bm X}_{t+1}, $
         \State Set  $\bm \delta=\bm X_{t+1}-\bm X_{t} $ \& $L=2$. 
        \Else 
        \State {Skip transmiting $ \bm X_{t}$ \& set $L=L+1$.} 
        \EndIf
 		\EndWhile
 	\end{algorithmic}
 \end{algorithm}
  Although in Algorithm \ref{pseudocode-CV}, the parameters of Online-MPLA are fixed at the beginning of the trip, the algorithm can be easily converted to an adaptive algorithm where the parameter setting can be frequently updated. Such an adaptive algorithm can assist in parameter updating that is needed by the operation center depending on the connected vehicle's location, the accuracy requirement for that location, and the number of connected vehicles in the vehicle's proximity. In case a request for parameter update is received from the operating center, the vehicle must terminate the current line segment, transmit the next coming data point, and continue Algorithm \ref{pseudocode-CV} with the updated setting. However, for the sake of simplicity, we present the algorithms assuming  fixed parameter setting throughout the trip.
  
As mentioned earlier, Algorithm \ref{pseudocode-CV} is written from the vehicle's standpoint. On the other end,  Algorithm \ref{pseudocode-Datacenter} summarizes  vehicle data approximation from the operation center standpoint. Whenever there is no transmitted data from the connected vehicle, the operation center will estimate the vehicle's information, knowing that the vehicle uses Algorithm \ref{pseudocode-CV} for efficient data transmission.
     \begin{algorithm}[H] 
     	\caption{Online-MPLA from operation center standpoint}\label{pseudocode-Datacenter}
     	\begin{algorithmic}[1]
     		\State Specify type of data that must be transmitted. 
     		\State Specify error threshold, $\bm \epsilon=(\epsilon^1,\cdots,\epsilon^d)$.
     		\State Specify maximum segment length, $K$.
     		\While {Trip has not ended}
     		\If {Receive data at time $t$}
     		\State $\tilde{\bm X}_{t}=\bm X_{t}$,
     		\State Set $\bm \delta=\tilde{\bm X}_{t}-\tilde{\bm X}_{t-1},$
     		\Else 
     		\State $\tilde{\bm X}_{t}$=$\tilde{\bm X}_{t-1}+\bm \delta$. 
     		\EndIf
     		\EndWhile
     		
     	\end{algorithmic}
     \end{algorithm}
     
 In the next section, we discuss  the numerical results from the empirical studies designed for the evaluation of applicability and accuracy  of the proposed methodology.  
     
\section{Empirical results}\label{sec:Exp}
In this section, we introduce the data set used in this work, discuss the experimental designs for the investigation, and report the numerical results in terms of accuracy, collection ratios, along with a comprehensive discussion. 
\subsection{Assessment of compression efficiency using CV data}
In this work, we have used the data provided by the Safety Pilot Model Deployment (SPMD) program, which was administered in Ann Arbor, Michigan \cite{DOTdata}. In order to evaluate the connected vehicle technologies under real-world condition, data from nearly 3000 vehicles with vehicle-to-vehicle (V2V) and vehicle-to-infrastructure (V2I) communication devices were collected during this program. In particular, the vehicles  in this program used dedicated short-range communications (DSRC) technology to transmit Basic Safety
Messages (BSMs) including vehicle operation information (e.g., speed, location, and heading) at a frequency of 10 messages per second \cite{bezzina2014safety}. 

From the SPMD data collected in April 2013, which is publicly available in \cite{DOTdata}, we randomly extracted data for 10,000 trips. Given the frequency of BSM data transmission, a trip is considered to be terminated if the gap between timestamps of two consecutive BSM data point for a specific vehicle ID is larger than 0.1 second. The extracted 10,000 trips are associated with  32,243,582 BSM data points, where each data point contains all the information regarding vehicle operation at a specific point in time. The 5th percentile, median, and 95th percentile of number of data points associated with these 10,000 trips are 200, 2076, and 9163, respectively. 

In this work, among the vehicle's information contained in each BSM data, we aim to efficiently collect and compress the speed and location, i.e. latitude and longitude, of the connected vehicles. However, it should be noted that the proposed methodology is readily applicable for any type of data requirement by the operation center. In what follows, we use data from the (randomly) selected 10,000 trips to evaluate and compare the performance of compressive  sampling, uniform sampling, and online-MPLA in efficient collection of connected vehicles data. 

This study concerns the effectiveness of collection strategies, and uses collection ratio and approximation error as comparison measures. Collection ratio is the ratio of the size of transmitted data to the size of actual data. Consider $\bm X^1$, $\bm X^2$, and $\bm X^3$ to include original speed, latitude, and longitude values for the selected 10,000 trips, respectively. Also let  $\tilde{\bm X}^1$, $\tilde{\bm X}^2$, and $\tilde{\bm X}^3$ to include approximated speed, latitude, and longitude values, respectively. To evaluate the accuracy of approximated speed, we report the median and $\ell_\infty$ norm of absolute error, $\bm e^s$, i.e., $ e^s_t=| \bm X^1_t-\tilde{\bm X}^1_t  |$. To evaluate the accuracy of approximated trajectory, we report the median  and $\ell_\infty$ norm  of the Euclidean distance between original and approximated locations, $\bm e^c$ i.e., $e^c_t=\sqrt{(\bm X^2_t-\tilde{\bm X^2_t})^2+(\bm X^3_t-\tilde{\bm X^3_t})^2}$.

\subsubsection{Data collection based on compressive sampling}
To collect data based on the compressive sampling approach, first a collection ratio must be fixed. To decide whether a data point must be transmitted or not, a random number between zero and one is generated. If the generated random number is smaller than the collection ratio, then the data point will be transmitted. The $\ell_1$ minimization in (\ref{eq:l1min}) must be solved to estimate the DCT coefficients, $\bm \alpha$, which is then used (\ref{eq:idct1}) to approximate the original data. In \cite{lin2018efficient}, to reduce the computation cost of $\ell_1$ minimization, the number of columns of $\bm \Psi$ is set to be 200. In other words, the signal associated with a vehicle trip data is divided into smaller signals (with the length of 200) and is recovered by solving the $\ell_1$ minimization problem for each sub-signal. We use the same approach to make our results comparable with the results reported in \cite{lin2018efficient}.

\begin{figure} [h]
	\centering 	
		\includegraphics[width=0.7\linewidth]{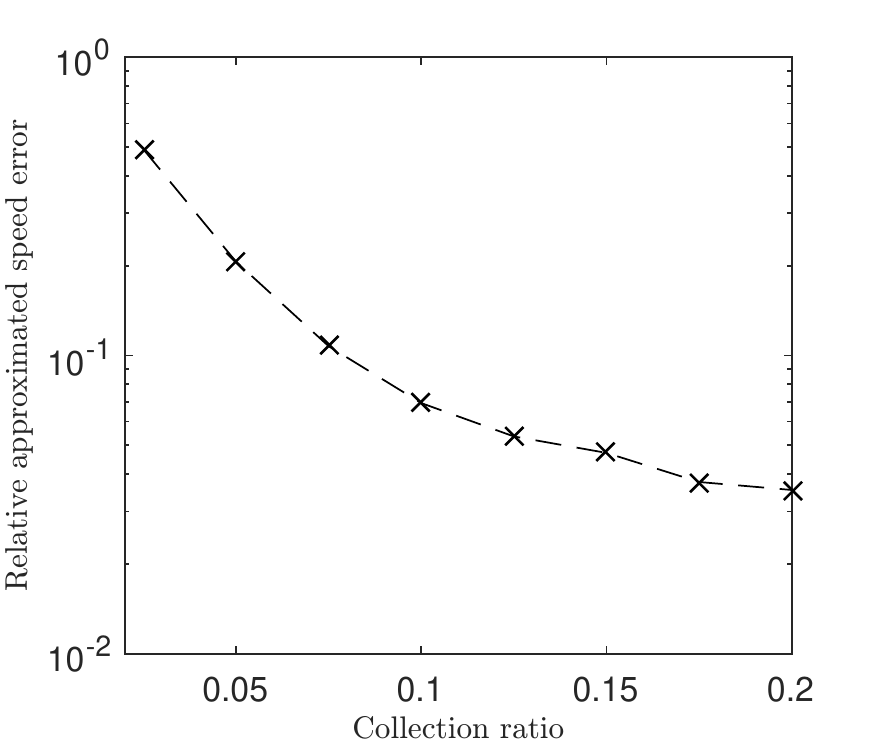}	
		\caption{Relative $\ell_2$ error for recovered speed using compressive sampling.}
			\label{fig:speed-cs-l2}	
\end{figure}

In \cite{lin2018efficient}, the relative $\ell_2$ error for recovered speed, i.e. ${\left \| \bm X^1-\tilde{\bm X}^1 \right \|_2}/{\left \| \bm X^1 \right \|_2}$, is used to evaluate the accuracy of the compressive sampling approach. Figure \ref{fig:speed-cs-l2} shows the relative $\ell_2$ error for the recovered speed, calculated at different collection ratios. Similar to the results in \cite{lin2018efficient}, we observe that a relative error smaller than 0.05 can be achieved for collection ratios larger than 0.15. 

In order to further evaluate the performance of this approach, in Figure  \ref{fig:speed-cs} we show the median and $\ell_\infty$ norm of the absolute error for recovered speed data using compressive sampling. It can be seen that although the  median of error is small, the $\ell_\infty$ norm of the error can be significantly large. Figure  \ref{fig:trajectory-cs} shows the median and $\ell_\infty$ norm of Euclidean distance between the original  trajectory and that approximated using compressive sampling. Again, it can be seen that the $\ell_\infty$ norm of the error can be significantly large. Based on these results it can be concluded that the relative $\ell_2$ error alone cannot be an appropriate measure for the accuracy offered by compressive sampling, in particular  because it does not offer any guarantee on maximum approximation error.

\subsubsection{Data collection based on uniform sampling}
In uniform sampling, every $i$th data point is transferred, where $i$ is an integer whose value  is the inverse of the collection ratio.   Once the signal is sampled at these equal intervals, a simple linear interpolation can be used to approximate the signal.

\begin{figure} [h]
	\centering 	

	\begin{subfigure}[t]{0.48\linewidth}
		\includegraphics[width=1\linewidth]{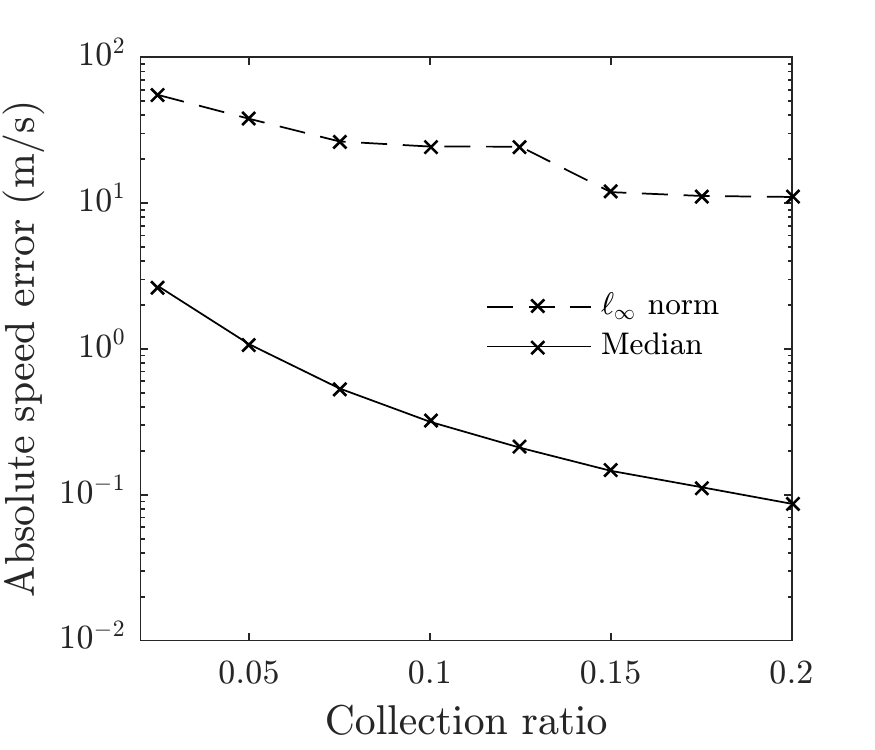}
		\caption{\footnotesize{Speed approximation accuracy using compressive sampling}}
		\label{fig:speed-cs}		
	\end{subfigure}
	\begin{subfigure}[t]{0.48\linewidth}
	\includegraphics[width=1\linewidth]{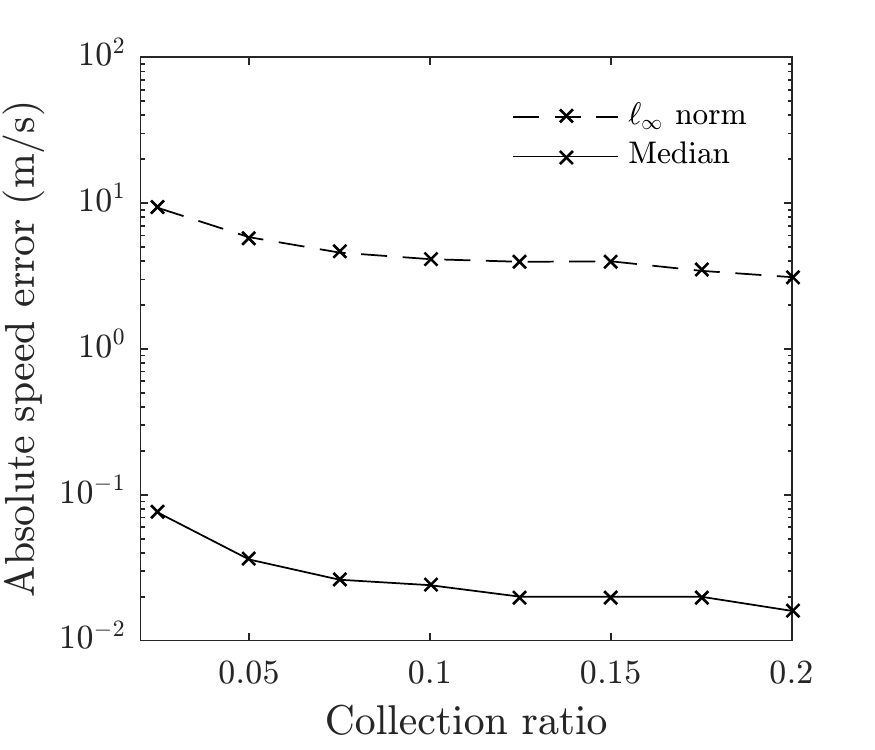}
	\caption{\footnotesize{Speed approximation accuracy using uniform sampling}}
	\label{fig:speed-uniform}		
\end{subfigure}
	\\
	\begin{subfigure}[t]{0.48\linewidth}
		\includegraphics[width=1\linewidth]{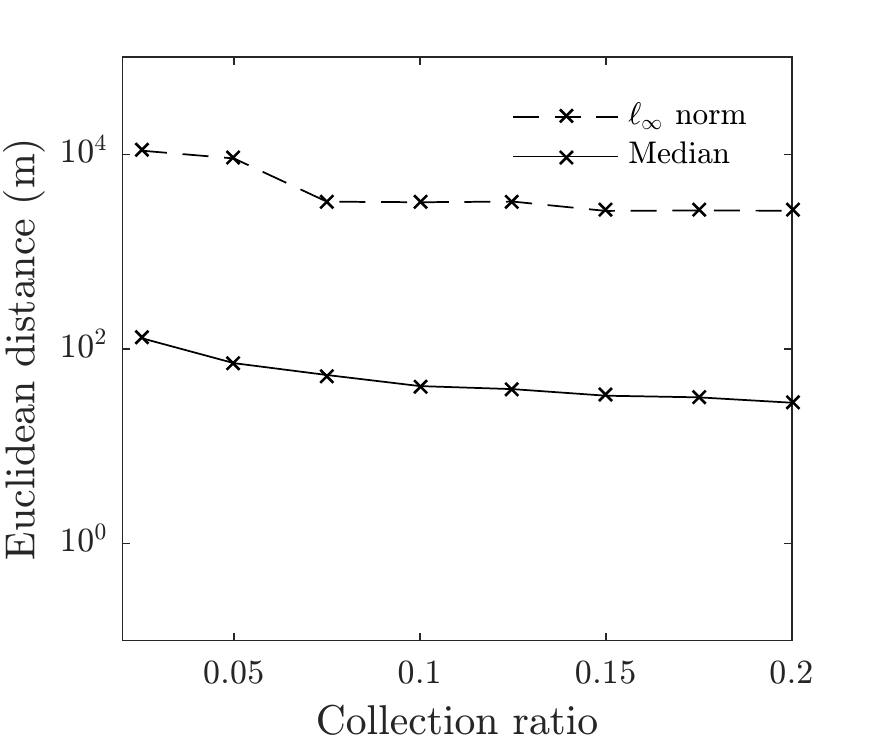}
		\caption{\footnotesize{Trajectory approximation accuracy using compressive sampling}}
		\label{fig:trajectory-cs}
	\end{subfigure}
		\begin{subfigure}[t]{0.48\linewidth}
			\includegraphics[width=1\linewidth]{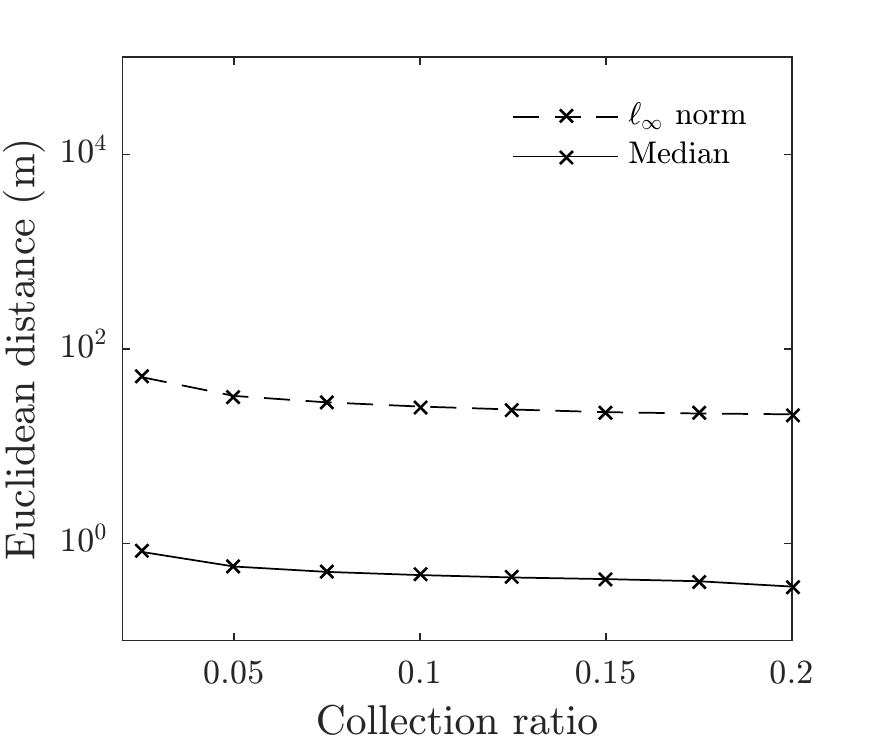}
				\caption{\footnotesize{Trajectory approximation accuracy using uniform sampling}}
			\label{fig:trajectory-uniform}
		\end{subfigure}
	
	\caption{Approximation accuracy when data is collected based on uniform and compressive sampling approaches. }
	\label{fig.uniform_CS}
\end{figure}

Figure \ref{fig:speed-uniform} shows the median and $\ell_\infty$ norm of the approximated speed absolute error. It can be seen that for each collection ratio, the approximation error using uniform sampling is substantially smaller compared to compressive sampling. Figure \ref{fig:trajectory-uniform} shows the relatively small median and $\ell_\infty$ norm of the Euclidean distance between the original and approximated trajectory.

Similar to compressive sampling, uniform sampling does not provide any guarantee on the maximum absolute error, either. However, the maximum absolute errors for speed and trajectory are restricted by physics laws. For example, the maximum Euclidean distance cannot be larger than the potential maximum speed of a vehicle multiplied by the elapsed time between two consecutive sampling points. Consequently, the maximum error obtained by uniform sampling is substantially smaller compared to compressive sampling. It can be seen that when collection ratio is 0.2, i.e. the elapsed time between two sampling points is 0.5 s, the maximum Euclidean distance is about 20 m and $10^3$ m using uniform sampling and compressive sampling, respectively. This confirms the advantage of uniform sampling over compressive sampling when absolute error measure is considered. 

Depending on the kind of application that uses  connected vehicle data, the accuracy offered by uniform sampling might be satisfactory or not. In Figure \ref{fig.uniform_CS}, it can be seen that at the collection ratio of 0.05, the maximum absolute error for speed and trajectory estimates are 6 m/s and 30 m, respectively. Considering the very small median of error, a collection ratio of 0.05 can provide  accurate travel time estimates when several connected vehicles are present in a road segment. However, the travel time estimation may not be accurate if only a few connected vehicles travel on a road segment. Besides, if the connected vehicle is also required to transmit information that are binary (e.g. whether there is a blockage in the road), such information can be easily missed in uniform sampling. Also, in spite of the ongoing debate about the future use of connected vehicle data in law enforcement \cite{Angelo2018, Keith2018}, researchers can enable platforms for such usages, the most immediate of which is the forensics analysis performed after  accidents. In such cases, high accuracy and guaranteed maximum approximation error is vital for a reliable forensics assessment. These considerations  call for an efficient data collection approach that guarantees the precision.   

\subsubsection{Efficient data collection using Online-MLPA}
Algorithm \ref{pseudocode-CV} by construction requires the specification of  maximum error thresholds for speed and trajectory approximations. It then  guarantees that the approximation errors never exceed the specified thresholds. This is because the algorithm skips collecting a data point only if the approximation errors are within the respective thresholds.

In order to perform a rather comprehensive study, we consider 16 different scenarios based on different precision thresholds for the three measured quantities, that is the speed, longitude and latitude. Table \ref{tab:LPA} shows the collection ratio associated with each threshold scenario. It should be noted that a 0.0001 degree difference in the coordinate  roughly corresponds to a 11.1 meter difference. Therefore, when the threshold for longitude and latitude is set to be 0.0001 degree the maximum Euclidean distance will not exceed 15.7 m.   
\begin{table}[h]
	\caption{Collection ratio achieved by using online-MLPA for efficient data collection.  }
	\scalebox{0.9}{	
		\begin{tabular}{|c|c|c|c|c|}
			\hline
			\multirow{2}{*}{\begin{tabular}[c]{@{}c@{}}Threshold\\ scenario\end{tabular}}& \multicolumn{3}{c|}{Specified collection error threshold} & \multirow{2}{*}{\begin{tabular}[c]{@{}c@{}}Collection\\ ratio\end{tabular}} \\ \cline{2-4}
			& \begin{tabular}[c]{@{}c@{}}$\epsilon^1$\\ (m/s)\end{tabular} & \begin{tabular}[c]{@{}c@{}}$\epsilon^2$\\ (degree)\end{tabular} & \begin{tabular}[c]{@{}c@{}}$\epsilon^3$\\ (degree)\end{tabular} &  \\ \hline
			1 & 0.5 & $0.5 \times 10^{-4}$ & $0.5 \times 10^{-4}$ & 0.181 \\ \hline
			2 & 1 & $0.5 \times 10^{-4}$ & $0.5 \times 10^{-4}$ & 0.151 \\ \hline
			3 & 1.5 & $0.5 \times 10^{-4}$ & $0.5 \times 10^{-4}$ & 0.142 \\ \hline
			4 & 2 & $0.5 \times 10^{-4}$ & $0.5 \times 10^{-4}$ & 0.137 \\ \hline
			5 & 0.5 & $1 \times 10^{-4}$ & $1 \times 10^{-4}$ & 0.130 \\ \hline
			6 & 1 & $1 \times 10^{-4}$ & $1 \times 10^{-4}$ & 0.096 \\ \hline
			7 & 1.5 & $1 \times 10^{-4}$ & $1 \times 10^{-4}$ & 0.085 \\ \hline
			8 & 2 & $1 \times 10^{-4}$ & $1 \times 10^{-4}$ & 0.079 \\ \hline
			9 & 0.5 & $1.5 \times 10^{-4}$ & $1.5 \times 10^{-4}$ & 0.114 \\ \hline
			10 & 1 & $1.5 \times 10^{-4}$ & $1.5 \times 10^{-4}$. & 0.078 \\ \hline
			11 & 1.5 & $1.5 \times 10^{-4}$ & $1.5 \times 10^{-4}$ & 0.067 \\ \hline
			12 & 2 & $1.5 \times 10^{-4}$ & $1.5 \times 10^{-4}$ & 0.061 \\ \hline
			13 & 0.5 & $2 \times 10^{-4}$ & $2 \times 10^{-4}$ & 0.106 \\ \hline
			14 & 1 & $2 \times 10^{-4}$ & $2 \times 10^{-4}$ & 0.070 \\ \hline
			15 & 1.5 & $2 \times 10^{-4}$ & $2 \times 10^{-4}$ & 0.057 \\ \hline
			16 & 2 & $2 \times 10^{-4}$ & $2 \times 10^{-4}$ & 0.052 \\ \hline
		\end{tabular} \label{tab:LPA}}
\end{table}

It can be seen that the achieved collection ratios are very small. For instance, to guarantee that speed approximation error is always smaller than 1.5 m/s and the location error is always smaller than 31.4 m, i.e. the thresholds for longitude and latitude are 0.0002 degree, the online-MLPA algorithm needs to collect about 5\% of the vehicle data.  This is while the uniform sampling approach requires 20\% of the vehicle data to achieve the same maximum error for the speed estimate; and compressive sampling results in significantly larger maximum error even at 20\% collection ratio.  

\begin{figure} [h]
	\centering 	
	\includegraphics[width=0.7\linewidth]{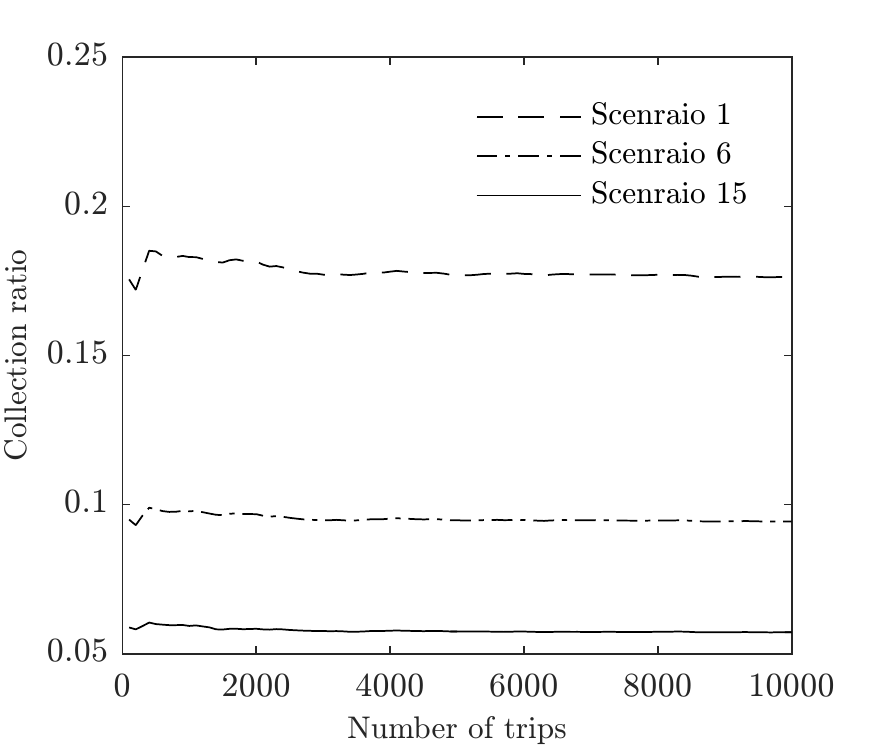}	
	\caption{Average collection ratio vs. the number of trips}
	\label{fig:Conv}	
\end{figure}

The  collection ratios reported in  Table \ref{tab:LPA} are averaged over the 10,000 trips studied in this work. To assure that 10,000 trips provide accurate estimation of collection ratio, we have conducted a convergence study. A representative convergence result  is depicted in Figure \ref{fig:Conv} where the change in average collection ratios due to the change in the number of trips is shown to have converged for three representative threshold scenarios. This also means that Table \ref{tab:LPA}  can be used when the operator needs to achieve a pre-specified collection ratio (due to e.g. limited communication) and seeks to identify the appropriate error thresholds to be set in the online-MLPA algorithm. 

\begin{figure} [h]
	\centering 	
		
		\begin{subfigure}[t]{0.48\linewidth}
			\includegraphics[width=1\linewidth]{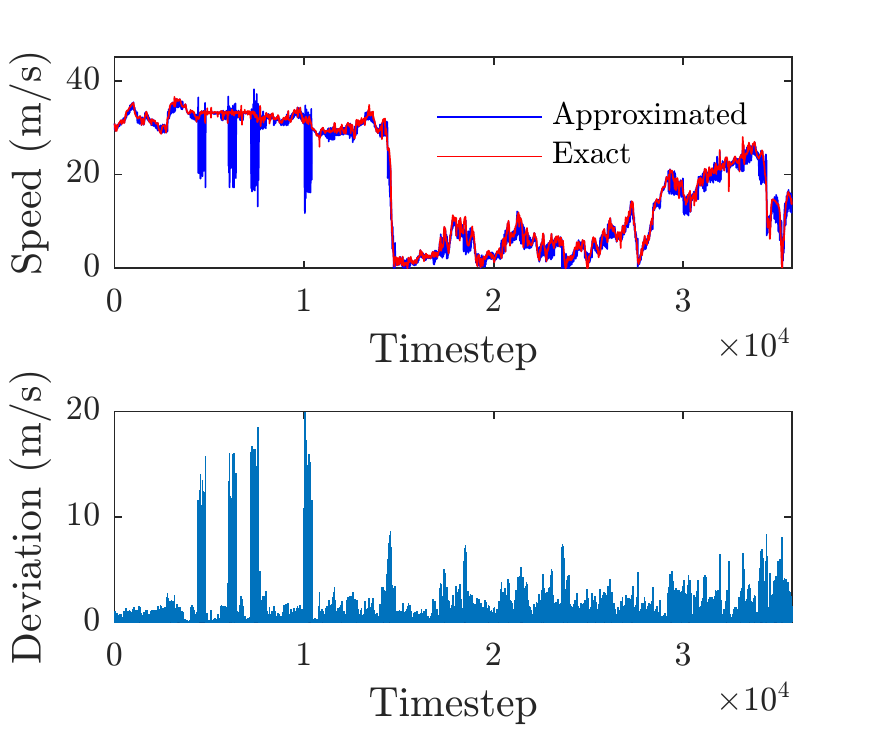}
			\caption{\footnotesize{Compressive sampling}}
			\label{fig:signal-Cs}		
		\end{subfigure}
		\begin{subfigure}[t]{0.48\linewidth}
			\includegraphics[width=1\linewidth]{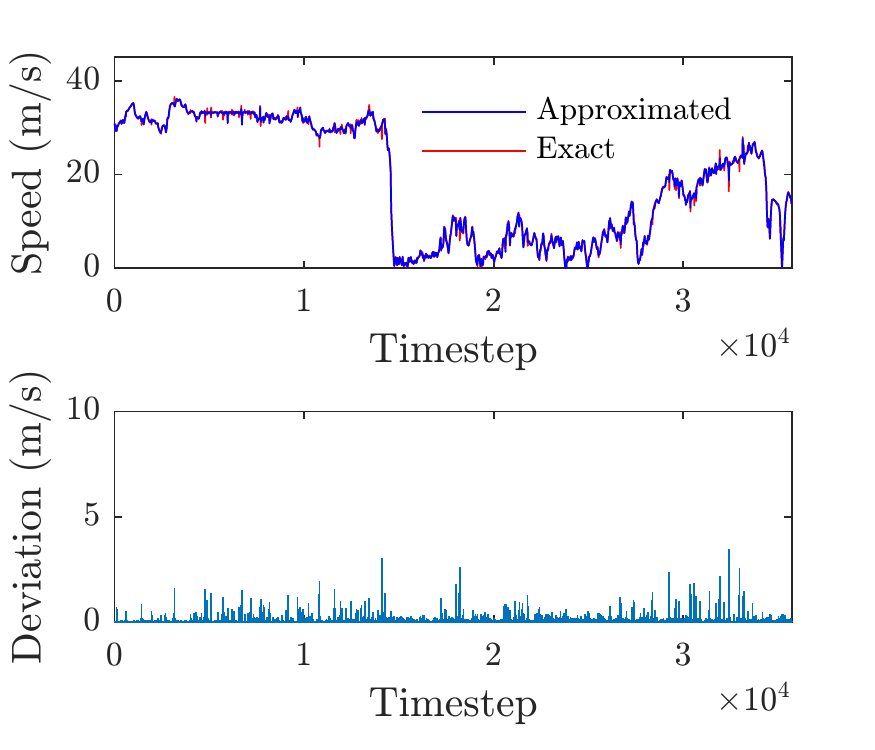}
			\caption{\footnotesize{Uniform sampling}}
			\label{fig:signal-Un}		
		\end{subfigure}
		\\
		\begin{subfigure}[t]{0.48\linewidth}
			\includegraphics[width=1\linewidth]{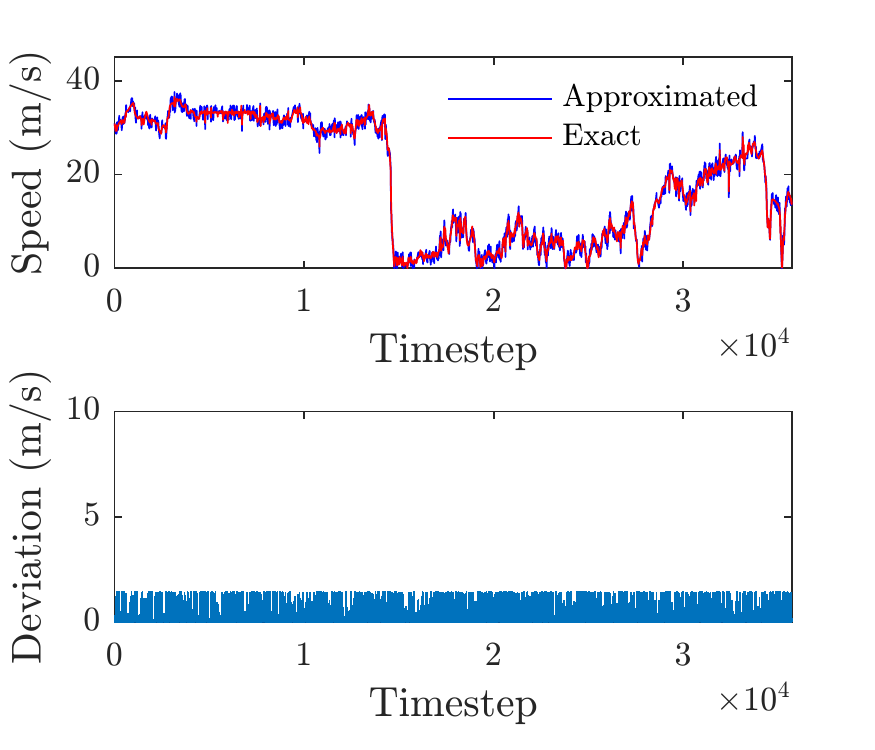}
			\caption{\footnotesize{Online-MPLA}}
			\label{fig:signal-MLPA}
		\end{subfigure}
		
	\caption{Original and approximated speed (top) and absolute values of deviations (bottom) for a single trip using (a) compressive sampling (b) uniform sampling (c) Online-MLPA. For online-MPLA, error thresholds are set the same as scenario 15 in Table \ref{tab:LPA}. Collection ratio is fixed to be 0.063 for uniform and compressive sampling. }
	\label{fig:speed}	
\end{figure}

 Next, let us fix the error thresholds  to be 1.5 m/s for speed and 0.0002 degrees for latitude and longitude for online-MLPA. To visualize the quality of a representative approximation obtained by Algorithm \ref{pseudocode-CV} and comparing it with uniform and compressive sampling, the exact and approximated values for the speed using the three approaches for a particular trip are shown in Figure \ref{fig:speed}. This trip was made by the vehicle with assigned ID `2176' on 04/10/2013, and included 35,723 samples points. Using Algorithm \ref{pseudocode-CV}, only 2,256 sample points were  transmitted. Using these transmitted points, Algorithm \ref{pseudocode-Datacenter} approximated the speed signal at the original 35,723 time steps. In Figure \ref{fig:signal-MLPA}, it can be seen that the approximation is sufficiently accurate and always follows the specified error threshold of 1.5 m/s. For a fair comparison between MLPA and uniform and compressive sampling, the compression ratio for these two sampling approaches was fix to be 0.063, which is the compression ratio achieved by online-MPLA. As can be seen in these figures, several instances of large deviations from exact speed values are observed in both uniform sampling and compressive sampling approximations. This highlights the necessity of an efficient data collection approach that restricts the maximum approximation error. 
      
\begin{figure} [H]
	\centering 	
	\includegraphics[width=0.7\linewidth]{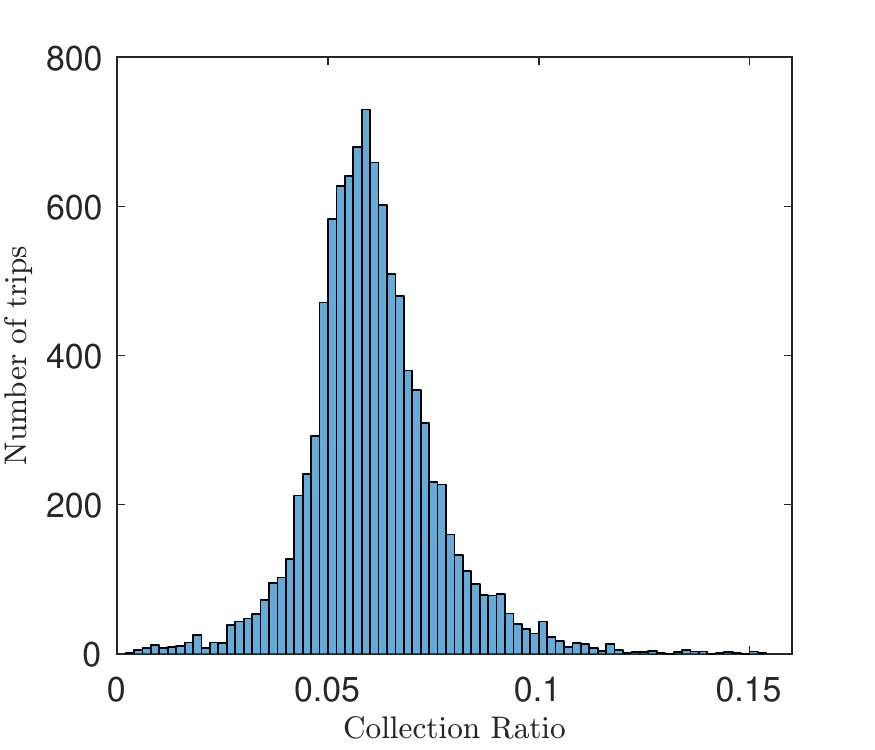}	
	\caption{Histogram of collection ratio achieved by online-MLPA for 10000 trips. Error thresholds are set the same as scenario 15 in Table \ref{tab:LPA}.}
	\label{fig:CRhist}	
\end{figure}

Figure \ref{fig:CRhist} shows the variation in the collection ratios calculated for the 10,000 trips.  It can be seen that the maximum observed collection ratio is 0.16. But we also observed that only 1.5\% of trips have a collection ratio larger than 0.1. Moreover, the collection ratio can be as small as 0.01. This variability in the collection ratios is due to the  different change patterns in the actual vehicle data. It can be seen that by not fixing the collection ratio, we can avoid unnecessary data collection when vehicle data doesn't involve frequent significant changes. 

\begin{figure} [h]
	\centering 	
	\includegraphics[width=0.7\linewidth]{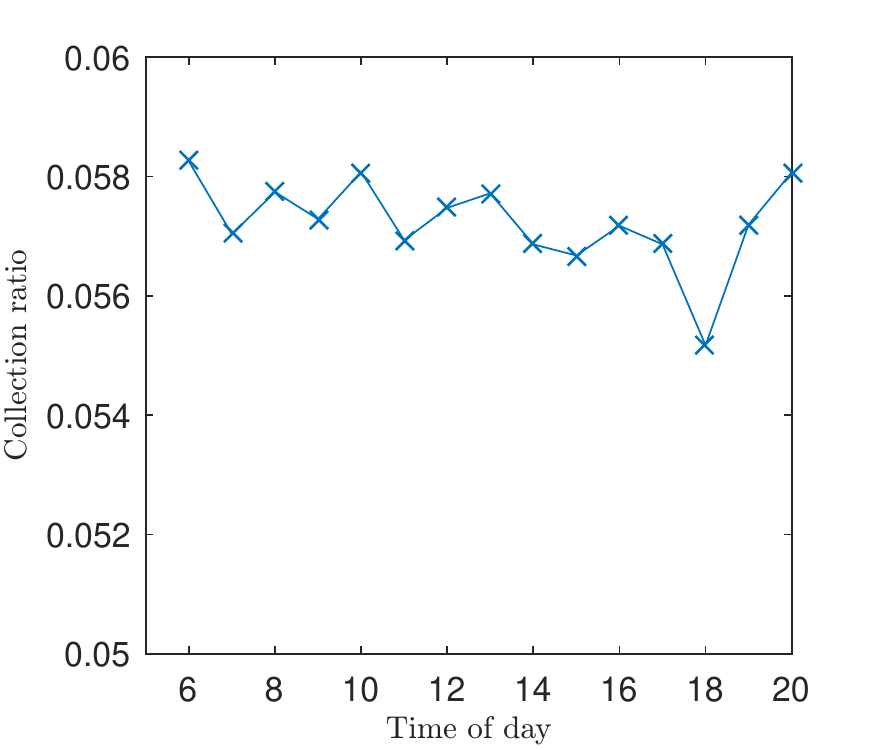}	
	\caption{Collection ratio  achieved by online-MLPA versus the time of day. Error thresholds are set the same as Scenario 15 in Table \ref{tab:LPA}.}
	\label{fig:CRvstime}	
\end{figure}

Figure \ref{fig:CRvstime} shows the average collection ratio for different times of day. It can be seen that collection ratio has a very small variation within a day. This numerically validates the efficiency of online-MLPA when used at any time of the day. 

\subsection{Study of travel time estimation using simulation data}
\begin{figure} [h]
	\centering 	
	\includegraphics[width=1\linewidth]{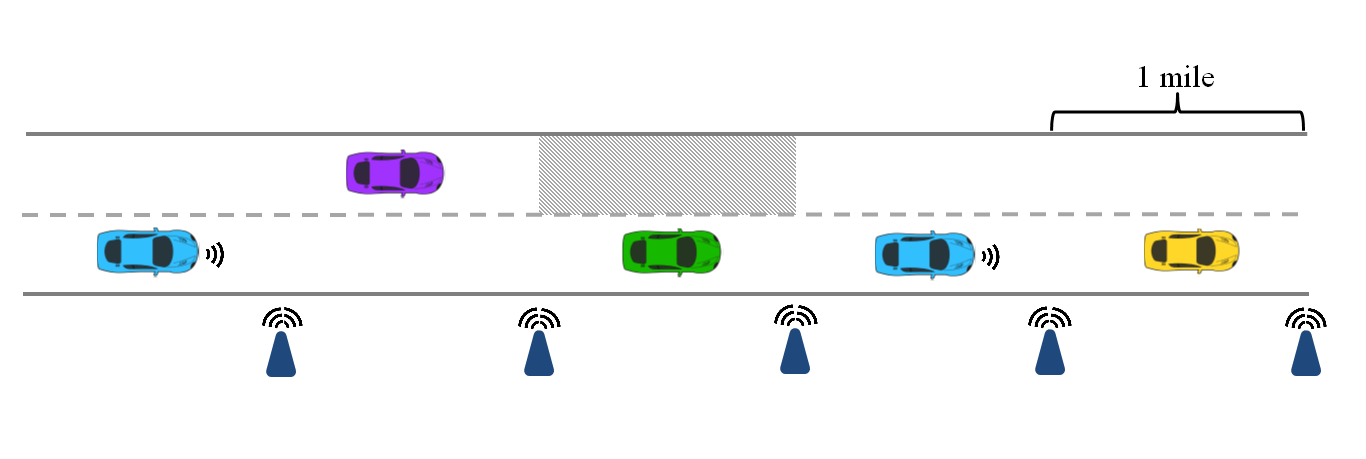}	
	\caption{Schematic view of the five-mile two-lane freeway section used in the simulation study.}
	\label{fig:freeway}	
\end{figure}

This section includes results from a simulation study, that is performed to evaluate the accuracy of travel time estimation given different data collection approaches. In particular, we used SUMO \cite{behrisch2011sumo}, which is an open-source microscopic simulation software, to simulate a five-mile two-lane freeway section. Figure \ref{fig:freeway} shows a schematic view of the freeway section, where one roadside unit (RSU) that has transceivers is considered at the end of each mile. RSUs receive data from on board units (OBUs) of connected vehicles and then send data to transportation management centers for real-time system analysis.

\begin{figure} [h]
	\centering 	
	\includegraphics[width=0.8\linewidth]{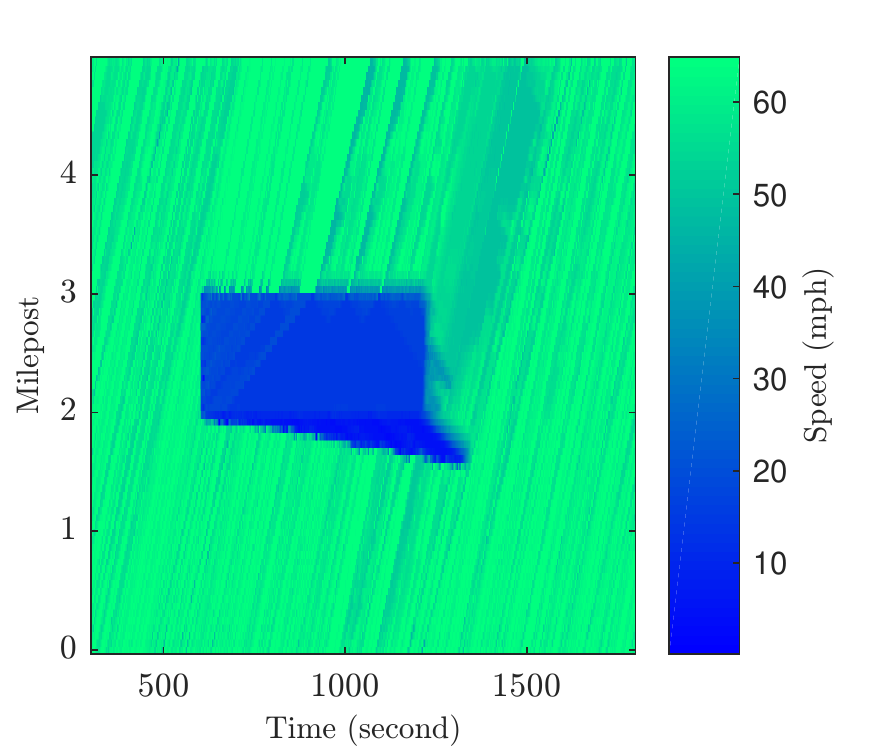}	
	\caption{Time-space-speed diagram for the simulated freeway section. In the section between mileposts 2 and 3, one of the lane is closed and the speed limit of the other one is reduced between the times 600 s and 1200 s.}
	\label{fig:Time-space}	
\end{figure}

The total simulation time was considered to be 1,800 seconds and the demand for the simulation period was set to be 1,100 vehicles. The speed limit for freeway section is enforced at 65 mph. For simplicity of explanation, let us divide the simulation time into six sub-periods, $t_1,t_2, \cdots, t_6 $, each indicating five minutes of simulation time. Similarly, let us divide freeway section into five sub-sections, $s_1,\cdots, s_5 $, each indicating one mile of freeway section. In order to impose congestion condition, one lane of $s_3$ was closed and the speed limit of the other lane was reduced to 20 mph during $t_3$ and $t_4$. Figure \ref{fig:Time-space} shows the time-space diagram for traffic flow speed, where the impact of lane closure and reduced speed limit in $s_3$ is evident. It can be seen that a shockwave was created that caused congestion in $s_2$ as well as $s_3$.  The $t_1$ period is considered as warm-up period and is excluded form Figure \ref{fig:Time-space} and further analysis.  

Our end goal in this section is to estimate travel time solely using data collected from connected vehicles. We assume that connected vehicles transmit all the data stored in their OBUs once they pass an RSU, without any loss. It is assumed that OBUs have a limited capacity, and that if the capacity of an OBU is reached before a vehicle passes an RSU and have the chance to transmit the data, earliest records of data will be deleted to make room for new data points. This data deletion can potentially lead to loss of important traffic information, something an efficient collection of data can prevent and effectively enable more accurate travel time estimation. In the following section, the  accuracy of travel time estimations given different data collection methods are compared.     

\subsubsection{Travel time estimation accuracy}
To evaluate the travel time estimation accuracy, we first need to establish travel time estimation approach. We calculate the exact travel time using the trajectory of all vehicles in the simulation. Let $T_j^i$ denote the exact travel time of section   $s_i$, during period $t_j$, where $i\in \{1,\cdots, 5 \}$, and $j \in \{2, \cdots, 6\}$. We then consider $T_j^i$   to be the average travel time of vehicles which passed $s_i$ during $t_j$. To estimate the travel time from connected vehicle speed data, Equation (\ref{eq:avspeed}) is used  to calculate $\bar{v}_j^i$, which is the average speed of $s_i$ during $t_j$
\begin{equation}\label{eq:avspeed}
\bar{v}_j^i= \frac{1}{N_j} \sum_{t=1}^{N_j}\left(\frac{\sum_{c=1}^{C^i_t} v^c_t}{C^i_t}\right),
\end{equation}
where $N_j$ is the total number of time steps in period $t_j$, $C^i_t$ is the total number of connected vehicle in $s_i$ at time $t$, and $v^c_t$ is the speed of connected vehicle $c$ at time $t$.  Given $\bar{v}_j^i$, travel time of $s_i$ during  $t_j$ can be estimated as
\begin{equation}\label{eq:avTT}
\tilde{T_j^i}= \frac{l_{i}}{\bar{v}_j^i},
\end{equation}
where $l_{i}$ is the length of $s_i$. The relative error of estimated travel time is then simply calculated as 
\begin{equation}
e_{r}= \frac{1}{25}\sum_{j=2}^{6} \sum_{i=1}^{5} \frac{|\tilde{T_j^i}-{T_j^i}|}{{T_j^i}}.
\end{equation}

Connected vehicles are considered to record data every 0.1~s, in order to be consistent with the SPMD data. We fix the maximum error thresholds for online-MLPA to be the same as scenario 16 in Table \ref{tab:LPA}. After using these threshold, the resulting collection ratio was found to be 0.083. For a fair comparison of travel time estimation accuracy using different data collection methods, the collection ratios for uniform and compressive sampling was also set to be 0.083.  It can be seen that achieved collection ratio is larger than the collection ratio in Table \ref{tab:LPA} for scenario 16, that was obtained from real world CV data. This is because in the simulation, we have set up an extreme situation such that the majority of simulated vehicles pass through an enforced congestion. 

\begin{figure} [h]
	\centering 	
	\includegraphics[width=0.7\linewidth]{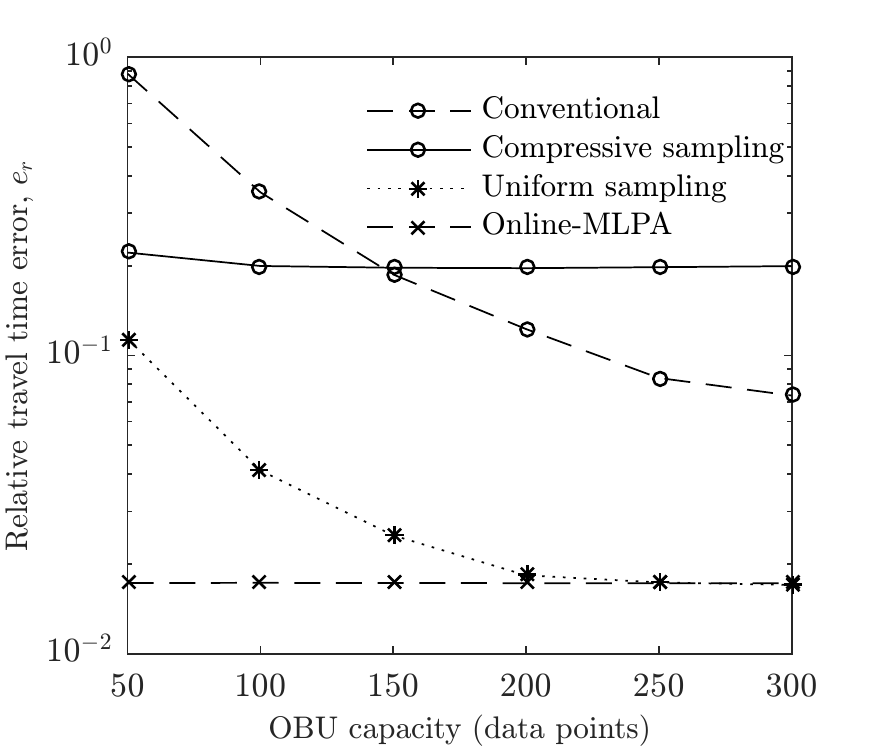}	
	\caption{Relative error of travel time estimation made by the four  collection methods versus  the capacity of OBUs.  }
	\label{fig:TTfixCV}	
\end{figure}

Let us first fix the percentage of connected vehicles to be 50\%. Figure \ref{fig:TTfixCV} shows the estimated travel time relative error for different data collection methods versus  OBU capacity. The capacity of OBU determines how many recordings of data points, each  including timestamps, speed, and location, can be stored. In \cite{kianfar2013placement} the maximum number of data points that could be stored in OBU is considered to be 30. In \cite{lin2018compressive}, OBU capacity was considered to be between 50 and 300 data points. In this figure, the conventional method refers to the case where all data points are recorded, and as can be seen, this approach yields extremely large travel time estimation errors when OBUs have very low capacity. This is because using conventional approach only most recent data points are stored and the communicated data has limited spatio-temporal coverage. It can also be seen that compressive sampling approach results in large travel time estimation error, which does not decrease by increasing the OBU capacity. This is a result of poor approximation accuracy of compressive sampling when collection ratio is smaller than 0.1 (see Figure \ref{fig:speed-cs-l2}). As expected, uniform sampling results in significantly better accuracy compared to compressive sampling. However, online-MLPA outperforms uniform sampling and results in less than 2\% relative error, even for OBUs with very low capacity. This is because the communicated data has larger spatio-temporal coverage, since it was collected only when a change in (speed/location) pattern happens, rather than following a fixed rate.  

\begin{figure} [h]
	\centering 	
	\includegraphics[width=0.7\linewidth]{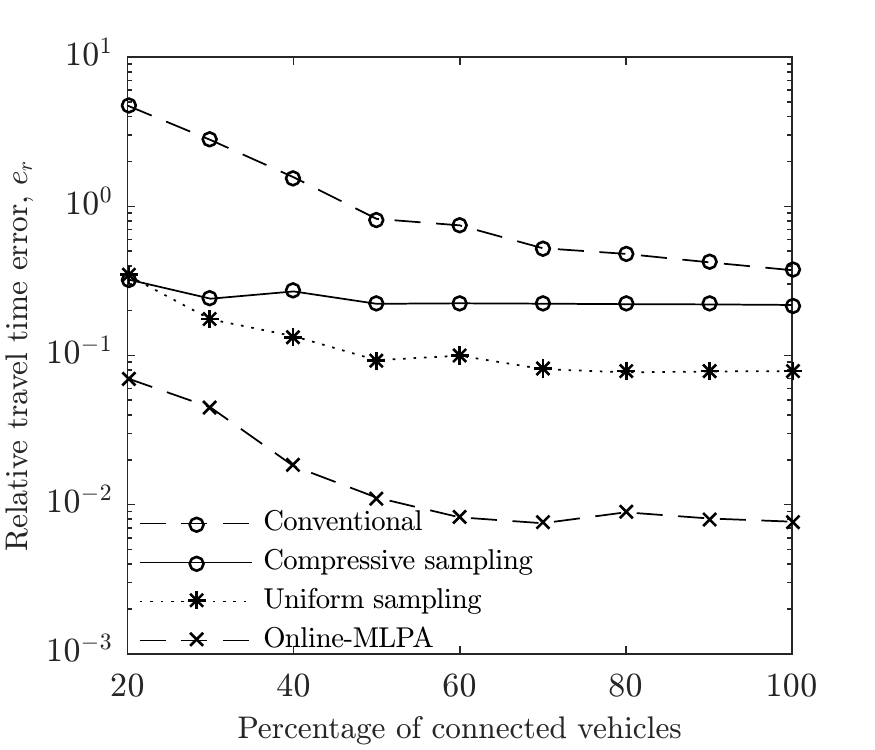}	
	\caption{Relative error of travel time estimation made by the four  collection methods versus the percentage of connected vehicles. }
	\label{fig:TTfixOBU}	
\end{figure}

To evaluate how different penetration level of connected vehicle will impact the accuracy of travel time estimation, we fix the OBUs capacity to be 50 data points and evaluate the accuracy versus the percentage of connected vehicles. As it can be seen in Figure \ref{fig:TTfixOBU}, similar accuracy is achieved for different percentages of connected vehicles when compressive sampling approach is used. This is because the error is mainly due to the poor approximation  that  compressive sampling yields at small collection ratios regardless of the percentage of connected vehicles. For uniform sampling and online-MLPA approaches, travel time estimation accuracy remains roughly the same for penetration percentages larger than 50\%. It can also be seen that online-MLPA results in the best accuracy for all the percentage levels. This is because online-MLPA  provides larger spatio-temporal coverage compared to all the other methods.  
\section{Conclusion}\label{sec:Conclusion}
In this work, an online multidimensional linear approximation approach was used to efficiently collect connected vehicle data given a required precision level. Real world connected vehicle data available in Safety Pilot Model Deployment dataset was used to evaluate the efficiency of proposed approach. It was shown that using the proposed method only a small subset of data must be collected  in order to comply even with  a very strict  precision requirement. Furthermore, a simulation study was performed to evaluate the accuracy of travel time estimation using the  compressed data. It was shown that the proposed compression method produces  travel time estimates that are more accurate than all the competing alternatives.  As a future direction, one can investigate the application of the proposed method for vehicle to vehicle communication. The challenge would be establishing precision requirements that do not compromise safety while allowing efficient communications.   

\section*{Acknowledgment}
This material is based in part upon work supported by the 
National Science Foundation under Grant No.  CMMI-1752302.

\bibliographystyle{IEEEtran}

\bibliography{IEEEabrv,bibfile}
\begin{IEEEbiographynophoto}{Negin Alemazkoor}
	is a PhD candidate in Sustainable and Resilient
	Infrastructure Systems program in the Department of
	Civil and Environmental Engineering at the University
	of Illinois at Urbana-Champaign. She received her
	Bachelor's degree in Civil Engineering from Sharif
	University of Technology and her Master's from Texas
	A\&M University. Her research interests include uncertainty quantification, data-driven and stochastic modeling, reliability and sensitivity analysis for infrastructure systems. 
\end{IEEEbiographynophoto}
\begin{IEEEbiographynophoto}{Hadi Meidani}
	earned his Ph.D. in Civil Engineering, together with a M.S. in Electrical Engineering  from the University of Southern California (2012).  After Ph.D. graduation, he was a Postdoctoral Scholar in the Department of Aerospace and Mechanical Engineering at USC (2012-2013) and a Postdoctoral Research Associate in the Scientific Computing and Imaging Institute at the University of Utah (2013-2014). Since Fall 2014, he has been an Assistant Professor in the Department of Civil and Environmental Engineering at the University of Illinois at Urbana-Champaign. In 2018, he received an NSF CAREER Award  to create advanced computational methods for the analysis and design of interdependent infrastructure systems.
	
\end{IEEEbiographynophoto}

\end{document}